\documentclass[12pt,preprint]{aastex}

\def\geqsim{\lower.73ex\hbox{$\sim$}\llap{\raise.4ex\hbox{$>$}}$\,$}
\def\leqsim{\lower.73ex\hbox{$\sim$}\llap{\raise.4ex\hbox{$<$}}$\,$}

\newcommand{\myemail}{aross2@uiuc.edu}

\newcommand{\SDSSpixt}{{\em SDSSpix~}}

\slugcomment{{\apj  Accepted}}
\shorttitle{N-Order Angular LRG Correlations in the SDSS }
\shortauthors{Ross et al.}

\begin{document}

\title{Normalization of the Matter Power Spectrum via Higher-Order Angular Correlations of Luminous Red Galaxies }

\author{Ashley J. Ross\altaffilmark{1}, Robert J. Brunner\altaffilmark{1,2},  Adam D. Myers\altaffilmark{1}}

\email{\myemail}

\altaffiltext{1}{Department of Astronomy,Ê 
University of Illinois at Urbana-Champaign,Ê 
Urbana, IL 61801}
\altaffiltext{2}{National Center for Supercomputing Applications,
Champaign, IL 61820}

\begin{abstract}
 We present a novel technique to measure $\sigma_8$, by measuring the dependence of the second-order bias of a density field on $\sigma_8$ using two separate techniques.  Each technique employs area-averaged angular correlation functions ($\bar{\omega}_N$), one relying on the shape of $\bar{\omega}_2$, the other relying on the amplitude of $s_3$ ($s_3 =\bar{\omega}_3/\bar{\omega}_2^2$).  We confirm the validity of the method by testing it on a mock catalog drawn from Millennium Simulation data and finding $\sigma_8^{measured}- \sigma_8^{true} = -0.002 \pm 0.062$.   We create a catalog of photometrically selected LRGs from SDSS DR5 and separate it into three distinct data sets by photometric redshift, with median redshifts of 0.47, 0.53, and 0.61.  Measurements of $c_2$, and $\sigma_8$ are made for each data set, assuming flat geometry and WMAP3 best-fit priors on $\Omega_m$, $h$, and $\Gamma$.  We find, with increasing redshfit, $c_2 = 0.09 \pm 0.04$, $0.09 \pm 0.05$, and $0.09 \pm 0.03$ and $\sigma_8 = 0.78 \pm 0.08$, $0.80 \pm 0.09$, and $0.80 \pm 0.09$.  We combine these three consistent $\sigma_8$ measurements to produce the result $\sigma_8 = 0.79 \pm 0.05$.  Allowing the parameters $\Omega_m$, $h$, and $\Gamma$ to vary within their WMAP3 1$\sigma$ error, we find that the best-fit $\sigma_8$ does not change by more than 8$\%$ and we are thus confident our measurement is accurate to within 10$\%$.  We anticipate that future surveys, such as Pan-STARRS, DES, and LSST, will be able to employ this method to measure $\sigma_8$ to great precision, and will serve as an important check, complementary, on the values determined via more established methods.
 \end{abstract}


\section{Introduction}
The normalization of the matter power spectrum is parameterized as the {\it rms} mass fluctuation within a top-hat radius of 8$h^{-1}$Mpc and denoted $\sigma_8$.  Measuring the amplitudes of the matter power spectrum, and thus $\sigma_8$, is complicated by the fact that most of the matter in the Universe is dark and that we must therefore rely on ``tracers" of the matter --- in most cases galaxies.  There is no guarantee, of course, that galaxies will cluster in the same manner as dark matter.  The relationship between the clustering of galaxies and dark matter is known as the ``bias" (see, e.g., \citealt{Kais}).  The bias essentially shifts the amplitudes of the galaxy power spectrum relative to the matter power spectrum and there is thus a strong degeneracy between the bias and $\sigma_8$.  Therefore, precisely determining $\sigma_8$ is important, as until this is accomplished, the full form of the relationship between the clustering of dark matter and the clustering of galaxies will remain ambiguous.

Measurements of $\sigma_8$ made using cluster counting techniques find mixed results.  Using the X-ray temperature and luminosity functions and fully marginalizing over the cluster scaling relation, \cite{Pier} found $\sigma_8 = 0.77^{+0.05}_{-0.04}$, while \cite{Henry} found $\sigma_8 = 0.66 \pm 0.16$.  Using a self-calibration technique and the red-sequence to optically identify clusters in the red-sequence cluster survey, \cite{Glad} found $\sigma_8 = 0.67^{+0.18}_{-0.13}$, while \cite{Rozo} found $\sigma_8 = 0.92 \pm 0.10$ using the SDSS maxBCG \citep{Koester} catalog.  Clearly, there is large variation in the measurements of $\sigma_8$ determined via cluster abundances.

Measurements made using data from the Wilkinson  Microwave Anisotropy Probe (WMAP) place some of the best constraints on the value of $\sigma_8$, yet still allow a wide range of values.  The first year WMAP (WMAP1) results found $\sigma_8 = 0.92 \pm 0.10$ \citep{WMAP1}, while the third year (WMAP3) results determined 0.744$^{+0.05}_{-0.06}$ \citep{WMAP3}.  The best-fit WMAP3 results vary significantly depending on the adopted constraints and priors.  In the currently accepted cosmological paradigm, inflation is a key ingredient.  When WMAP3 is constrained by a variety of inflationary models, the best-fit $\sigma_8$ are found to be as low as 0.702 $\pm 0.062$ \citep{WMAP3}.  If one instead combines the WMAP3 results with the Sloan Digital Sky Survey (SDSS) galaxy power spectrum \citep{Teg04}, $\sigma_8 = 0.772^{+0.041}_{-0.042}$.  Considering the importance of inflation to our understanding of the Universe, the constraints placed by WMAP3 are quite loose.  Analysis of the WMAP 5 year data (WMAP5; made public during the revision process of this work) yielded a best-fit 5-year mean value of 0.796 $\pm$ 0.036 \citep{WMAP5}.  Despite the precision of this value, if one considers the high and low values measured by WMAP1 and WMAP3 and the range in values determined via cluster counting techniques, the true value of $\sigma_8$ remains unclear.

In this paper, we present a novel technique to measure $\sigma_8$.  Our technique relies on the fact that the bias relationship may be non-linear and if the bias is non-linear, then both the amplitude and the shape of the correlation function (the Fourier transform of the power spectrum) are affected.  As the value of $\sigma_8$ affects only the amplitude of the correlation function measurement, the non-linear bias is not degenerate with $\sigma_8$.  Therefore, different techniques for measuring the extent of the non-linear bias can constrain not only the non-linear bias, but $\sigma_8$ as well.  

The bias relationship can be expressed as a Taylor expansion, with the parameters $b_1$ and $b_2$ representing the first- and second-order contributions (and $b_2$ thus being a measure of the non-linearity; see, e.g., \citealt{Gaz92}).  In general, it is convenient to express the second-order contribution as $c_2 = b_2/b_1$. The average overdensity increases as the scale becomes smaller, therefore the effect of $c_2$ increases on smaller scales.  If $c_2$ is positive, the correlation function will be increasingly amplified towards smaller scales.  Thus, transitioning between the non-linear and linear regime ($\sim$10 $h^{-1}$Mpc, the ``weakly non-linear regime") the bias relationship changes such that $b_2$ transitions from having little effect on the measurement to having an important effect.  Therefore, it is ideal to test $c_2$ using correlation functions at scales on either side of the weakly non-linear regime, as the effects of $b_1$ and $b_2$ can be decoupled by considering both small and large scales.

Photometrically selected luminous red galaxies (LRGs) from the SDSS DR5 imaging data are ideal for measuring correlation functions in the weakly non-linear regime.  The median redshift of SDSS LRGs is $\sim$0.52, meaning that 10 $h^{-1}$Mpc is equivalent to about 0.4 degrees---a scale at which angular galaxy correlation functions in SDSS can be calculated quite precisely.  Thus, by studying the shape of the 2-point area-averaged correlation function ($\bar{\omega}_2$) of LRGs around these scales, one can measure $c_2$ quite precisely.  This measurement is dependent on $\sigma_8$ and therefore one can determine the relationship between the non-linear bias of LRGs and $\sigma_8$.  A separate relationship between the non-linear bias and $\sigma_8$ can be found by utilizing the shape of the hierarchical amplitude $s_3 = \bar{\omega}_3/(\bar{\omega}_2)^2$ (see, e.g., \citealt{R07}; R07 from hereon).  These two relationships are nearly orthogonal, thereby allowing unambiguous determinations of $b_1$, $c_2$, and $\sigma_8$.  

This is not the first study of the clustering of photometrically selected LRGs.  \cite{Pad06} and \cite{Blake} selected LRGs from SDSS imaging data and studied their clustering.  These efforts measured the two-point correlation function and power-spectrum, respectively, at different redshifts to place tight constraints on the matter and baryon densities of the universe (but not $\sigma_8$).  Nor is this the first study to measure the higher-order clustering of LRGs, as \cite{K} measured the redshift space three-point correlation function of spectroscopically selected LRGs.  

Our work, however, has three major distinctions from these previous studies: (1) our data is drawn from the contiguous area of SDSS DR5, significantly improving our ability to calculate higher-order statistics at large scales, (2) we measure higher-order statistics using photometrically selected LRGs and thus the median redshift is higher than for any previous sample used to study the higher-order clustering of LRGs, and (3) we focus on determining the value of $\sigma_8$.  We will outline our methods for determining the two separate relationships between $\sigma_8$ and $c_2$ in \S2.  In \S3, we will describe the creation of a photometric redshift catalog of LRGs drawn from the SDSS DR5 imaging data, which follows the prescriptions of \cite{2slaq}.  In \S4, we will present precise measurements of the second- and third-order angular area-averaged correlation functions of LRGs.  We will display our measurements of the first- and second-order bias parameters of LRGs, and we will show that these bias parameters have a strong dependence on $\sigma_8$.  We combine these measurements to present a precise measurement of $\sigma_8$.

Unless otherwise noted, we assume a flat cosmology with parameters equal to the WMAP3 alone best-fit values ($\Omega_m$, $h$, $\Gamma$) $=$ (0.238, 0.73, 0.135), where $\Gamma$ is the shape parameter, and essentially parameterizes $\Omega_b$ assuming fixed $\Omega_m$ and $h$ (see, e.g., Equations 30 and 31 of \citealt{Eis98}).

\section{Methods}
\subsection{Angular Correlation Functions}\label{s:wn}
We estimate N-point area-averaged angular correlation functions, $\bar{\omega}_{N}$($\theta$), using a counts-in-cells technique identical to that used in R07.  This involves calculating the statistical moments of the over-densities contained in equal-area cells.  We create the cells using a modified version of the\SDSSpixt pixelization scheme originally developed by Tegmark, Xu, and Scranton\footnote{http://lahmu.phyast.pitt.edu/$\sim$scranton/SDSSPix/}.  The over-density for cell $i$ is defined as
 \begin{equation}
 \delta_{i} = \frac{\bar{n} - n_{i}}{\bar{n}}
 \end{equation}
\noindent where $\bar{n}$ is the average number of galaxies in a cell and $n_{i}$ is the number of galaxies in cell $i$.  The remaining details and equations required to determine $\bar{\omega}_{N}$($\theta$) are found in \cite{Ross}.  The hierarchical amplitudes are defined as
\begin{equation}
s_N = \frac{\bar{\omega}_N}{\bar{\omega}_2^{N-1}}
\end{equation}

\subsection{Bias and $\sigma_8$}
\label{sec:meth}
In order to determine the relationship between the bias parameter $c_2$ and $\sigma_8$, measurements of $\bar{\omega}_2$ and $s_3$ are compared to theoretical models constructed via matter power spectra.  In each case, the $z = 0$ value of $\sigma_8$ is input to the model, allowing $c_2(\sigma_8)$ to be calculated.

Model $\bar{\omega}_2$ are produced using power spectra calculated using the \cite{Smith} fitting formulae, as described in R07 \S6.  By using a modified version of Limber's equation \citep{Limb}, one can use the redshift distribution of our LRG catalog to invert the $P(k)$ and obtain $\bar{\omega}_2(\theta)$:
\begin{equation}
\bar{\omega}_2(\theta) = \frac{H_0\pi}{c}\int \int \left(\frac{{\rm d}n}{{\rm d}z}\right)^2\sqrt{\Omega_m(1+z)^3+\Omega_\Lambda} P(k,z)W^2_{2D}[\chi(z)\theta k]{\rm d}z{\rm d}k
\label{eq:w2}
\end{equation}
\noindent where $W_{2D} = 2\frac{J_{1}(x)}{x}$ is the top-hat two-dimensional window function, $\chi(z)$ is the comoving distance to redshift $z$, $P(k,z)$ is the matter power spectra, $k$ is the spectral index, and $J_1$ is the first-order Bessel function of the first kind, and this equation requires $\Omega_{total} = 1$ (see, e.g., \citealt{Ber02}).  This equation is then integrated using the assumed cosmology and the desired value of $\sigma_8$.

Both methods employed to measure $c_2$ are dependent on the product of $b_1$ and $\sigma_8$.  In order to account for this, we determine the first-order bias using the $\bar{\omega}_2$ measurement and the model $\bar{\omega}_2$ at $\sigma_8 = 0.8$ and denote it $b_{1,0.8}$.  This is calculated for scales where linear theory is a good approximation ($> 10h^{-1}$ Mpc).  Therefore, a valid expression for the first-order bias is given by $b_{1,LRG} = \left(0.8/\sigma_8\right)b_{1,0.8}$.  

The second-order bias can be determined by manipulating the overdensities in each cell used in the calculation of $\bar{\omega}_2$.  The overdensity of LRGs can be related to the overdensity of dark matter and bias terms via a Taylor expansion (to second order)
\begin{equation}
\delta_{LRG} = b_1\delta_{DM}+0.5b_2\delta^2_{DM}
\label{eq:bias}
\end{equation}
In order to apply a second order bias term to the LRG correlation measurement, one must solve Equation \ref{eq:bias} for $\delta_{DM}$, which, to second order, is
\begin{equation}
\delta_{DM}= \delta_{LRG}/b_1-2b_2\delta^2_{LRG}/b_1^3
\label{eq:bias2}
\end{equation}

Thus, in order to determine the first and second order bias of the LRGs, we use Equation \ref{eq:bias2} to apply a $b_1$ and a $b_2$ to each overdensity used in the measurement of $\bar{\omega}_2$ and match these altered measurements to the model $\bar{\omega}_2$.  To find the best fit $b_2$ for a given $\sigma_8$, one must simply calculate $\bar{\omega}_2$ for a sampling of $b_2$ values, calculate the $\chi^2$ for each, and minimize $\chi^2$ via iteration (fully accounting for covariance as noted in \S4).  This must then be repeated for all of the $\sigma_8$ values one wishes to test (which requires determining the model $\bar{\omega}_2$ for each $\sigma_8$).  This process requires the correlation functions to be calculated thousands of times, but is highly parallel.  In practice, we determine the $\chi^2$ values for selected $\sigma_8$/$b_2$ values on a grid of initial spacing 0.02 in $\sigma_8$ and 0.005 in $b_2$.  This grid is then refined in areas of rapidly changing $\chi^2$.  The $\chi^2$ value for any $\sigma_8$/$b_2$ pair is then found by using a two dimensional spline fit to the grid. This method will hereon be referred to as the {\it shape} method.

The other method we employ is detailed in R07, and takes advantage of the relationship \citep{Fry93}
\begin{equation}
 s_{3}=b_{1}^{-1}(s_{3,DM} + 3c_{2})
\label{eq:s3}
\end{equation}
where $s_{3}$ is the measured amplitude and $s_{3,DM}$ is the theoretical amplitude.  The $s_{3,DM}$ are determined at scales greater than 8 $h^{-1}$Mpc.  This is accomplished by calculating $\bar{\omega}_3$ via the integration of linear power spectra and redshift distributions given by \cite{Ber95},
\begin{equation}
\begin{array}{l}
\bar{\omega}_{3,DM} = 6\left(\frac{H_0\pi}{c}\right)^2\int\left(\frac{{\rm d}n}{{\rm d}z}\right) ^3\left[\Omega_m(1+z)^3+\Omega_\Lambda\right]{\rm d}z~\times\cr \\  \left\{\frac{6}{7}\left(\int kP(k)W^2_{2D}[D\theta k]{\rm d}k\right) ^2+ \nonumber\int kP(k)\left(W^2_{2D}[D\theta k]\right){\rm d}k\int k^2D\theta P(k)W_{2D}[D\theta k]W^{'}_{2D}[D\theta k]{\rm d}k\right\} 
\end{array}
\end{equation}
where $D$ is the comoving distance to the median redshift.  We calculate $\bar{\omega}_2$ using Equation \ref{eq:w2} and linear power spectra and then use $s_3 = \bar{\omega}_3/\bar{\omega}_2^2$.  By altering $\sigma_8$ and using the relationship described in \S 7.1, Equation 14, of R07, we can determine the 1$\sigma$ allowed region of $c_2$/$\sigma_8$.  We will refer to this approach as the {\it R07} method.  

\subsection{Testing via Mock Catalogs}
To test our new method, we took galaxies with $M_r <$ -23 from the Blaizot all-sky mock catalog created using the methods described in \cite{blaiz} and Millennium Simulation data \citep{mill}.  In order to select red galaxies we constrained the absolute magnitudes of our mock catalog to have $B-R >$ 1.4, as we discovered that the color of the simulated galaxies was bimodal about this value.  This yielded a sample of nearly 300,000 simulated LRGs.  These simulated LRGs had a median redshift of 0.2; this is significantly smaller than the LRGs to be used in our measurements, but nonetheless quite sufficient to test our measurement techniques.  

\begin{figure}
\plotone{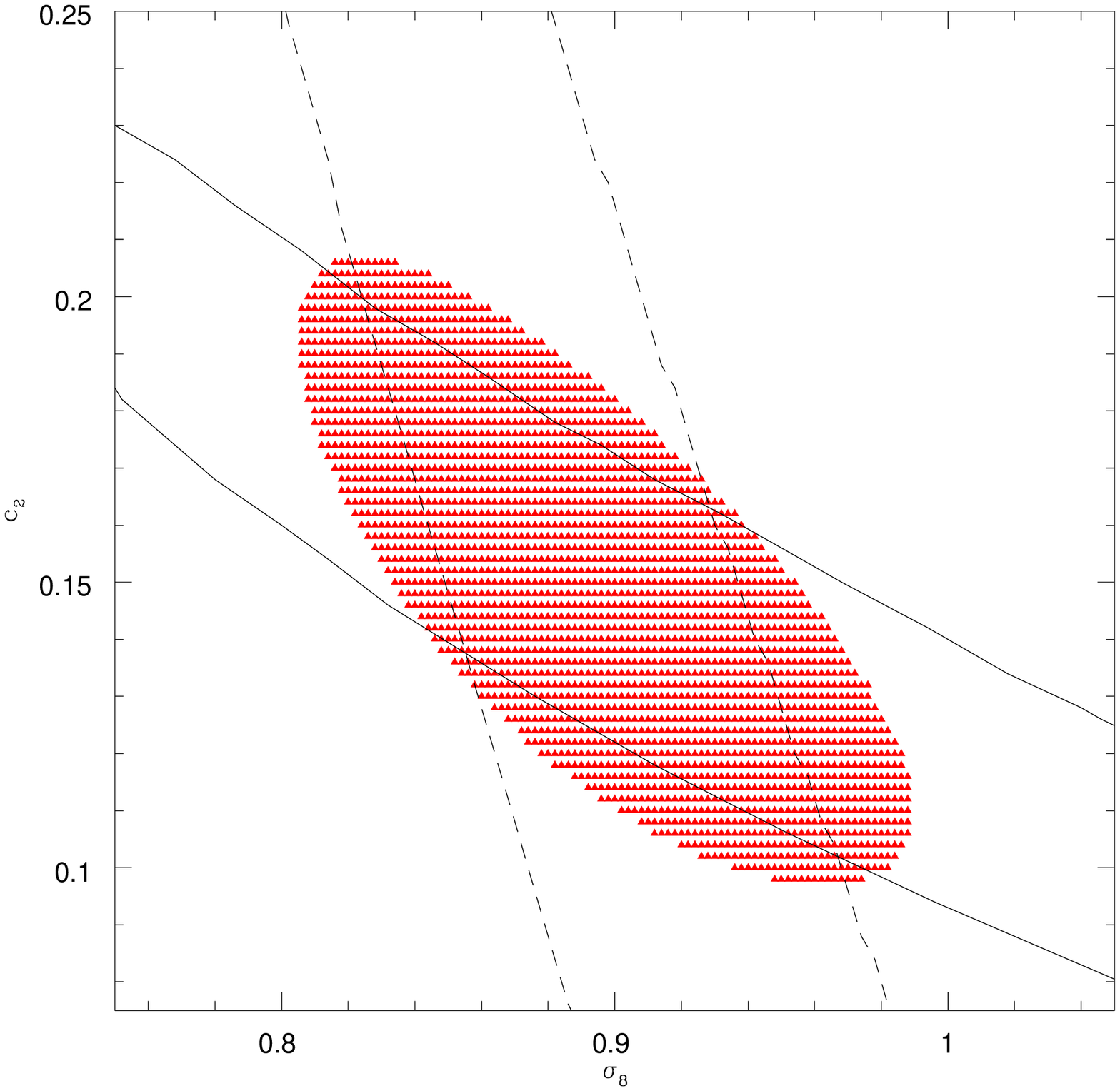}
\caption{The 1$\sigma$ allowed regions for $c_2 / \sigma_8$ determined by the shape method (solid black lines) and the R07 method (dashed black lines) for simulated LRGs from the Blaizot all-sky mock catalog \citep{blaiz} created using Millennium simulation data \citep{mill}.  These measurements are combined to produce the 1$\sigma$ (red) allowed region.  The 1$\sigma$ best-fit values are $c_2 = 0.15 \pm 0.04$ and $\sigma_8 = 0.898 \pm 0.062$, consistent with the Millennium input value of $\sigma_8 = 0.9$}
\label{fig:mill}
\end{figure}

Using our mock catalog, $\bar{\omega}_N$ and $s_N$ were calculated using the methods described in \S\ref{s:wn}.  In order to determine the bias of the simulated LRGs, we employed the methods described in \S2, to calculate model $\bar{\omega}_{N}$ and $s_N$.  This was done using $\sigma_8 = 0.8$ and the assumed cosmology of the Millennium Simulation (relevant parameters being $\Omega_m$, $h$, $\Gamma$ = 0.25, 0.73, 0.14; \citealt{mill}).  This allowed us to find $b_{1,0.8} = 2.04 \pm 0.02$, fit at scales $>$ 8.2 $h^{-1}$Mpc ($>$ 0.8$^{\circ}$).  With this value in hand, we could then use the {\it shape} method to find the $\chi^2$ values in the $b_2$/$\sigma_8$ parameter space.  This allowed us to find the best-fit $c_2$ value as a function of $\sigma_8$.  These 1$\sigma$ bounds, which represent $\Delta\chi^2 = 1$ from the minimum at that $\sigma_8$, are displayed on Figure \ref{fig:mill} by solid black lines.  For $\sigma_8 = 0.8$, we found $c_2 = 0.186 \pm 0.026$.

Using the {\it R07} method and setting $\sigma_8 = 0.8$, we found $c_2 = 0.41 \pm 0.09$.  The large disagreement with the value of $c_2$ determined by the {\it shape} and {\it R07} methods is expected for $\sigma_8 = 0.8$, as the two methods should agree only for the $\sigma_8$ used to create the mock catalog (0.9).  To find where the methods agreed, we calculated $\chi^2$ for the entire $c_2/\sigma_8$ parameter space using the {\it R07} method.  The resulting 1$\sigma$ allowed region, which represent $\Delta\chi^2 = 1$, is bounded by the dashed black lines in Figure \ref{fig:mill}.  Combining the $\chi^2$ distributions of the two methods produced the 1$\sigma$ ($\Delta\chi^2 = 2.3$ from the overall minimum) region displayed in red in Figure \ref{fig:mill}.  From these measurements, we determine $\sigma_8 = 0.898 \pm 0.062$ and $c_2 = 0.146 \pm 0.037$.  This measured $\sigma_8$ is entirely consistent with the input $\sigma_8 = 0.9$ of the Millennium Simulation.  This confirms that our method can indeed be used to measure the value of $\sigma_8$ both precisely and accurately. 

\section{Data}
We take data from the fifth data release (DR5) of the Sloan Digital Sky Survey.  To create a catalog of LRGs with photometric redshifts we applied the techniques described by \cite{2slaq} (abbreviated C07 from here on) applied to objects in the DR5 {\it PhotoPrimary} view.  Employing the color and magnitude cuts described by C07 produces a sample of just over 1.7 million objects.  As in C07, we found photometric redshifts by using the {\it annz} software \citep{F03} with the Two-Degree Field SDSS LRG and QSO (2SLAQ) spectroscopic LRG catalog \citep{Can06}, with stars removed, as training data.  In order to separate stars, we again used the {\it annz} software, and trained it on the 2SLAQ LRG target catalog.  In this case, we included the targeted objects determined to be stars and gave them a classification value of 0, while galaxies were given a classification value of 1.  This same method was employed by C07 to eliminate stars from their catalog.  

Our final catalog comprises only objects with classification values greater than 0.8.  Based on the training data, cutting on this value should reduce stellar contamination to less than 2\% while keeping 99.9\% of the LRGs.  This results in a catalog of 1,662,390 LRGs with a median photometric redshift of 0.52.  Our redshift distribution is nearly identical to the distribution found by C07.  These LRGs are then processed through the same imaging/reddening/seeing masks as in R07, leaving 1,168,702 objects.  We split these LRGs into three distinct photometric redshift ranges with similar numbers of objects, $0.4 < z < 0.5$ (444,175 LRGs), $0.5 < z < 0.57$ (398,250 LRGs), and $0.57 < z < 0.7$ (326,277 LRGs).  These data sets will be referred to as $Z_{0.47}$, $Z_{0.53}$, and $Z_{0.61}$, with median redshifts of 0.47, 0.53, and 0.61, respectively.  This gives us three distinct data sets that allow us to test the consistency of our measurements and that can be combined to increase the precision of our final $\sigma_8$ measurement.  

\section{Measurements}
We calculate the area-averaged angular correlation functions ($\bar{\omega}_N$) and hierarchical amplitudes ($s_N$) for photometrically classified SDSS DR5 LRGs using the methods described in \S\ref{s:wn}.  For every measurement, errors and covariance matrices are calculated using a jackknife method (e.g., \citealt{Scr02}), with inverse-variance weighting for both errors (e.g., \citealt{Mye05,Mye06}) and covariance (e.g., \citealt{Mye07}), identical to the one described in \S3.4 of R07.  This allows us to minimize $\chi^2$, fully accounting for covariance via
\begin{equation}
\chi^2 = \sum_{i,j}[\bar{\omega}(\theta_i) - \bar{\omega}_{m}(\theta_i)]C_{i,j}^{-1}[\bar{\omega}(\theta_j) - \bar{\omega}_{m}(\theta_j)]
\label{eq:Chi}
\end{equation}
where $C$ is the covariance matrix, and $i$ and $j$ refer to the $i^{th}$ and $j^{th}$ jackknife subsample.

\begin{figure}
\plotone{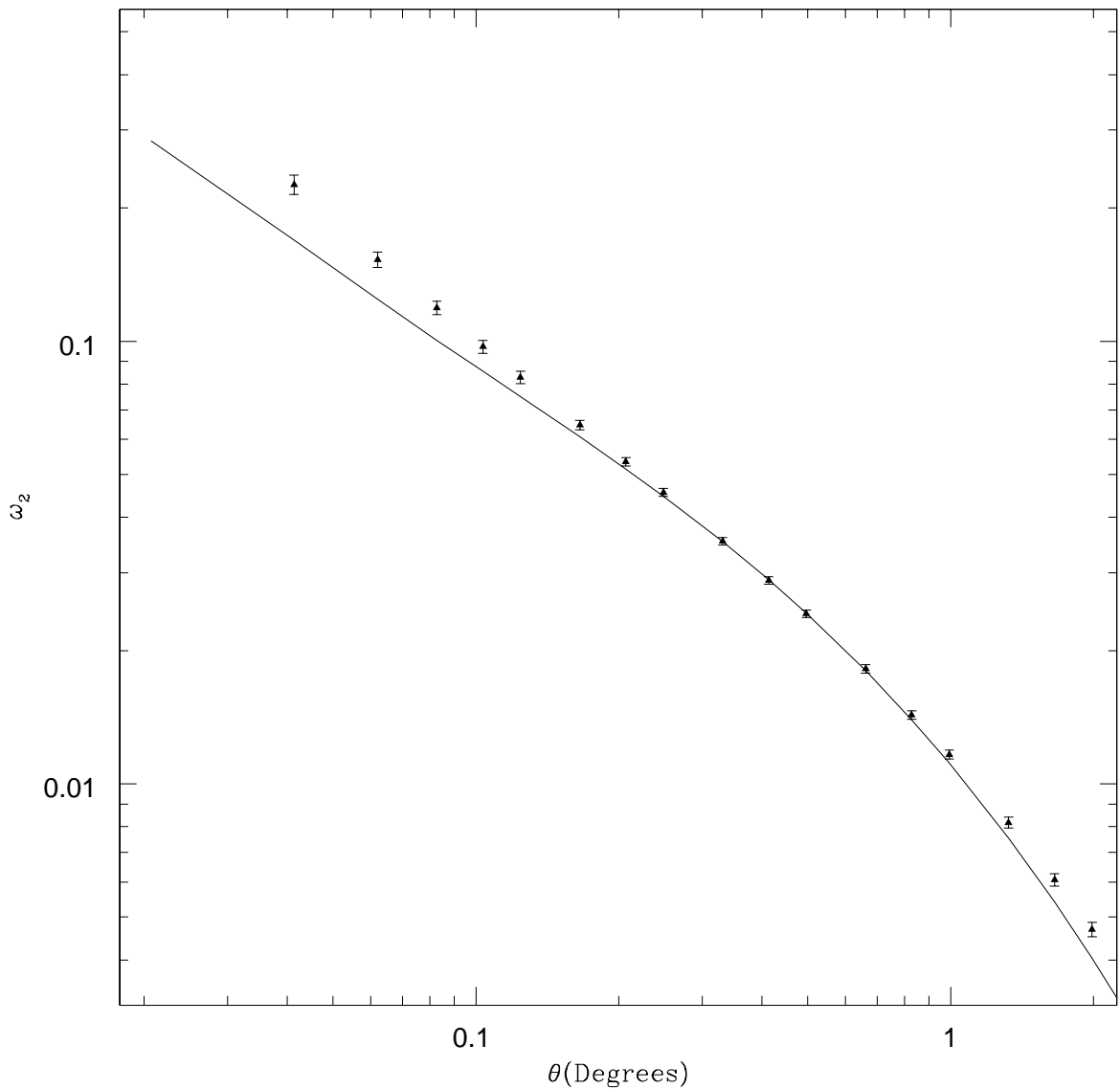}
\caption{The area-averaged angular two-point correlation function for photometrically-classified LRGs from SDSS DR5 with photometric redshifts $0.5 < z < 0.57$ ($Z_{0.53}$), divided by 1.63$^2$ (black triangles, $b_1 = 1.63$ for $\sigma_8 = 0.8$).  The solid line represents the model $\bar{\omega}_2$ calculated using \cite{Smith} power spectra and Limber's equation.  At scales smaller than 0.25$^\circ$, the measurement is larger than the model, suggesting positive second order bias.}
\label{fig:w2}
\end{figure}

Focusing first on $Z_{0.53}$, we fit $\bar{\omega}_2$ for measurements made between 0.4$^\circ$ and 1.6$^\circ$ (10.2 $h^{-1}$Mpc and 40.2 $h^{-1}$Mpc).  We determine $b_{1,0.8} = 1.63 \pm 0.02$.  The $\bar{\omega}_2$ is well fit by a single bias parameter in this range, as the $\chi^2 = 1.7$, $P(<\chi^2) = 0.89$.  Thus, we measure $b_{1,LRG} = 0.8/\sigma_8\times(1.63 \pm 0.02)$.  The $\bar{\omega}_2$ measurement is presented in Figure \ref{fig:w2}, divided by 1.63$^2$ (accounting for $b_{1,0.8}$) along with the model $\bar{\omega}_2$ at $\sigma_8 = 0.8$.  At scales less than 0.3$^\circ$, the measurement grows larger than the model, indicative of positive second-order bias.  The measurement also grows larger at scales greater than $\sim$2$^\circ$ (50.2 $h^{-1}$Mpc), but at these scales the errors begin to grow larger and systematics due to reddening and projection effects also increase.  Based on the results of \cite{Simon}, our theoretical curve, which employs a modified version of Limber's Equation, should not be accurate to better than 10$\%$ at scales greater than $\sim$2$^\circ$ for any of the redshift ranges we use.  Our measurement at 2$^{\circ}$ differs from the model by 8.5$\%$, thus the disagreement is no greater than would be predicted by \cite{Simon}.  We thus fit no measurements to scales greater than 1.6$^\circ$.  

Altering the $\bar{\omega}_2$ measurement using the {\it shape} method, and fitting the measurements at scales between 0.1$^\circ$ and 0.7$^\circ$ (equivalent to $\sim$ 2.5 to 17.6 $h^{-1}$ Mpc, there are 8 measurements in this range and thus 7 degrees of freedom), we measure $b_2 = 0.15 \pm 0.05$ for $\sigma_8 = 0.8$.  The fit is acceptable, as $\chi^2 = 0.60$, $P(<\chi^2) = 0.999$.  Attempting to fit the data with a single bias parameter model, we find $\chi^2 = 12.18$, $P(<\chi^2) = 0.09$.  For other redshift ranges, we find similar results.  Again for $\sigma_8 = 0.8$, we measure $b_{2, Z_{0.47}} =  0.150 \pm 0.040$ ($\chi^2 = 0.49$, $P(<\chi^2) = 0.999$).  A single bias parameter model is rejected at 87$\%$.  Finally, we find $b_{2, Z_{0.61}} = 0.165 \pm 0.025$ ($\chi^2$ = 1.91, $P(<\chi^2) = 0.96$).  A single bias parameter is rejected at $> 99\%$ for this redshift range.  Based on the marginal rejections of a single-parameter model, a two-parameter model is needed to fit the measurements for each redshift range.  For the two lower redshift ranges, the minimum $\chi^2$ values are quite small, implying that perhaps our error bars are overestimated for these redshift ranges, further implying our quoted errors on $b_2$ are overestimated.  

\begin{figure}
\plotone{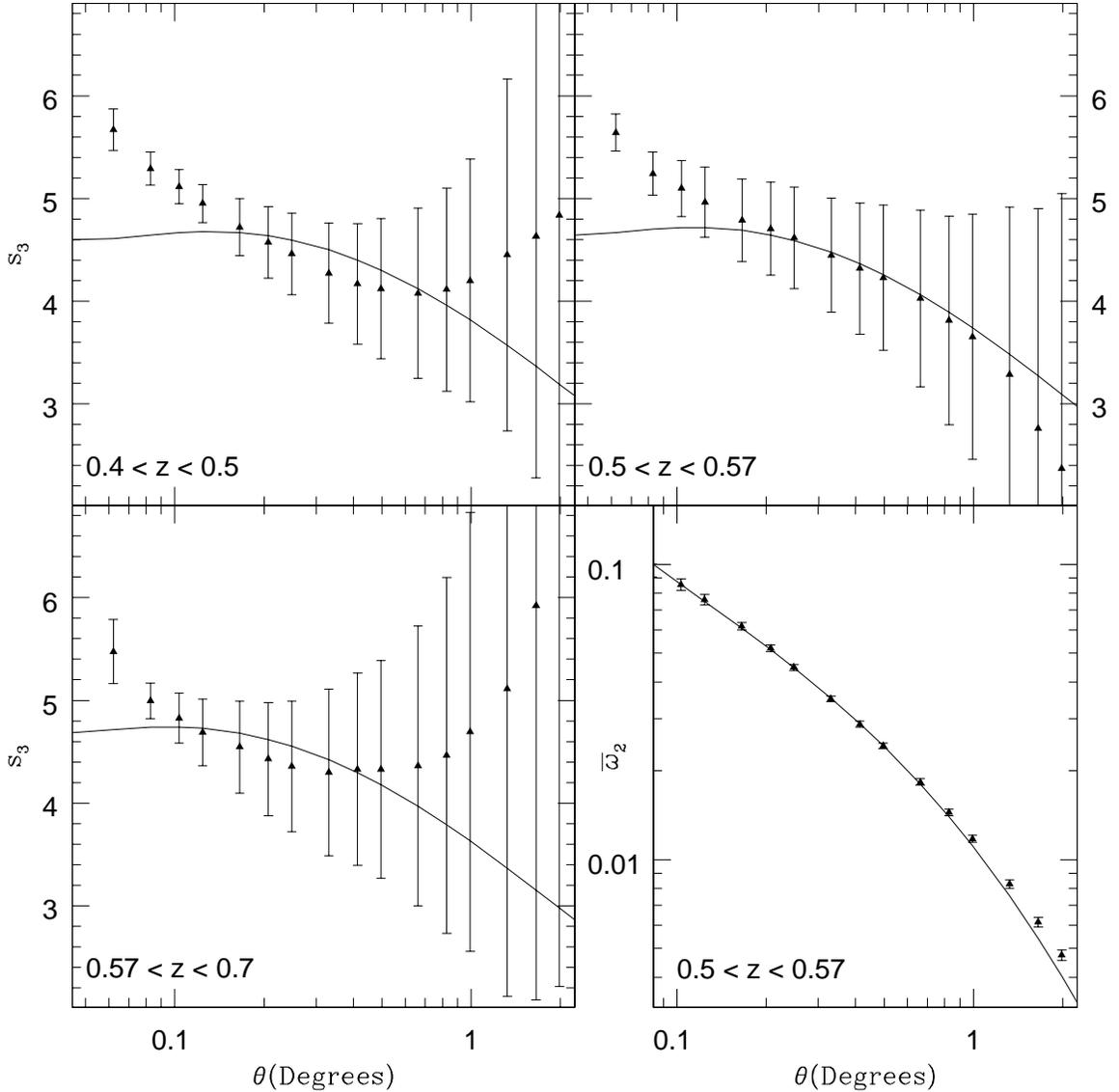}
\caption{Left to right, top to bottom, the first three panels display the measured $s_3$ (black triangles) for photometrically classified LRGs from SDSS DR5 with photometric redshifts $0.4 < z < 0.5$ ($Z_{0.47}$), $0.5 < z < 0.57$ ($Z_{0.53}$), and $0.57 < z < 0.7$ ($Z_{0.61}$).  In order to take the bias into account, $s_3$ is divided by the best-fit $b_1$ and then has three times the best-fit $c_2$ subtracted from it for each respective redshift range (for $\sigma_8 = 0.8$).  The solid line in each of these three panels is the model $s_3$ for $\sigma_8 = 0.8$.  The $Z_{0.53}$ data has a different shape than the other data sets at large scales, but due to the size of the error bars, all data sets are consistent with the model $s_3$ to at least 96$\%$.  The lower-right panel displays the bias-corrected measurement of $\bar{\omega}_2$ (black triangles) for $Z_{0.53}$ using the best-fit $b_1 = 1.63$, and $b_2 = 0.15$ for $\sigma_8 = 0.8$.  The solid black line is the model $\bar{\omega}_2$ for $\sigma_8 = 0.8$.  Visually, the fit appears to be as good as the $\chi^2 = 0.6$, $P(<\chi^2) = 0.998$ suggests.}
\label{fig:s3}
\end{figure}

In the bottom right panel of Figure \ref{fig:s3}, the $\bar{\omega}_2$ measurement corrected for $b_1 = 1.63$ and $b_2 = 0.15$ is displayed along with a theoretical $\bar{\omega}_2$ for $\sigma_8 = 0.8$.  The model curve clearly fits the data.  The other panels display the measured $s_3$ (black triangles) for $Z_{0.47}$, $Z_{0.53}$, and $Z_{0.61}$ (left to right, top to bottom) corrected for the best-fit $b_1$ and $c_2$ in accordance with Equation \ref{eq:s3}.  Each panel also includes a solid line displaying the model $s_3$.  The $Z_{0.53}$ measurement appears to be extremely consistent with the model, while the curves defined by the other two measurements do not share the same shape as the model.  Despite this fact, the size of the error bars allows the $Z_{0.47}$ and $Z_{0.61}$ measurements to appear consistent with the model. 

We use the $R07$ method to find $c_2$ for each data set.  For data sets $Z_{0.47}$ and $Z_{0.53}$ we fit between 0.4$^\circ$ and 1.6$^\circ$ (equivalent to 9.0 $h^{-1}$Mpc and 35.9 $h^{-1}$Mpc for $Z_{0.47}$ and 10.0 $h^{-1}$Mpc to 40.2 $h^{-1}$Mpc for $Z_{0.53}$; there are 7 measurements in this range and thus 6 degrees of freedom).  For $Z_{0.61}$, we fit between 0.3$^\circ$ and 1.6$^\circ$ (8.5 $h^{-1}$Mpc to 45.6 $h^{-1}$Mpc; again 7 measurements and 6 degrees of freedom).  We find that for $\sigma_8 = 0.8$, $c_{2, Z_{0.47}} = 0.08 \pm 0.12$, $c_{2, Z_{0.53}} = 0.07 \pm 0.13$ and $c_{2, Z_{0.61}} = 0.14 \pm 0.16$.  For $Z_{0.53}$, $\chi^2 = 0.033$, meaning $P(<\chi^2) = 1.0 - 1.0 \times 10^{-6}$.  This is a remarkably small $\chi^2$ value, which one might expect (to a degree) based on how well the measured values appear to match the model, despite the size of the error-bars.  This suggests that the error-bars are over-estimated for the $s_3$ measurements in this redshift range, and thus the error on our $c_2$ measurements may be overestimated as well.  For both the $Z_{0.47}$ and $Z_{0.61}$ ranges, $\chi^2 = 1.5$ with $P(<\chi^2) = 0.96$.  These values are quite reasonable, implying that if our errors are being overestimated, it is happening only for the $Z_{0.53}$ range.

\begin{figure}
\plotone{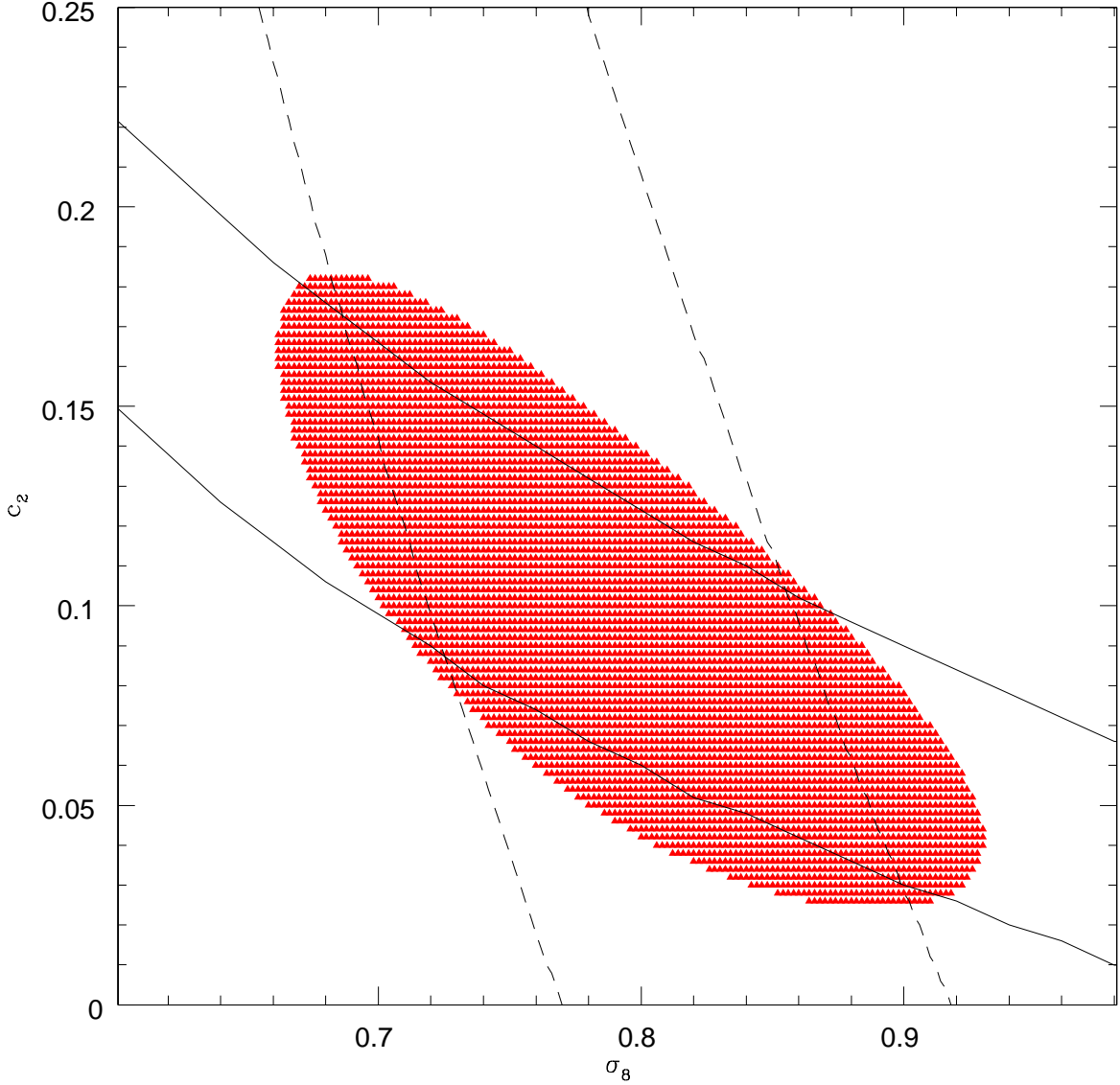}
\caption{The 1$\sigma$ allowed regions for $c_2 / \sigma_8$ determined by the {\it R07} method (dashed black lines) and the {\it shape} method (solid black lines) for photometrically selected LRGs from SDSS DR5 with $0.5 < z < 0.57$ ($Z_{0.53}$).  These measurements are combined to produce the 1$\sigma$ (red) allowed region.  The 1$\sigma$ best-fit values are $c_2 = 0.092 \pm 0.052$ and $\sigma_8 = 0.796 \pm 0.086$.}
\label{fig:contcom}
\end{figure}

Using the {\it shape} and {\it R07} methods (see \S \ref{sec:meth})  the $\chi^2$ values over the entire $\sigma_8$/$c_2$ parameter space are determined for each data sample.  The 1$\sigma$ allowed regions of $c_2$/$\sigma_8$ determined via the {\it R07} (solid black lines, $\Delta\chi^2 = 1$ from fixed $\sigma_8$) and the {\it shape} (dashed black lines, $\Delta\chi^2 = 1$ from fixed $\sigma_8$) methods are plotted in Figure \ref{fig:contcom} for $Z_{0.53}$.  Fortunately, the two methods bound significantly different regions of parameter space, allowing a precise determination of $c_2$ and $\sigma_8$.  Combining the two measurements produces the 1$\sigma$ (red triangles, $\Delta\chi^2 = 2.3$ from overall minimum) allowed regions for $c_2$/$\sigma_8$, and are also displayed in Figure \ref{fig:contcom}.  To 1$\sigma$ precision, we thus find $c_{2, Z_{0.53}} = 0.092 \pm 0.052$ and $\sigma_8 = 0.796 \pm 0.086$.  Repeating the process for $Z_{0.47}$, we find $c_{2, Z_{0.47}} = 0.088 \pm 0.041$ and $\sigma_8 = 0.776 \pm 0.080$ and for $Z_{0.61}$, we find  $c_{2, Z_{0.61}} = 0.092 \pm 0.033$ and $\sigma_8 = 0.798 \pm 0.094$.  

The best fit $\sigma_8$ for our three data sets are consistent to 0.275$\sigma$.  Combining the three measurements, we measure $\sigma_8 = 0.789 \pm 0.050$.  Adopting this value in order to determine the first order bias, we find $b_{1, Z_{0.47}} = 1.47 \pm 0.09$, $b_{1, Z_{0.53}} = 1.65 \pm 0.09$, and $b_{1, Z_{0.61}} = 1.80 \pm 0.10$.  These values make sense given that the median luminosity of the galaxies increases with redshift since our sample is not volume limited.  If we multiply each of the best-fit $c_2$ measurements by 0.789 divided by the best-fit $\sigma_8$ for each respective data set (approximately correct for small changes in $\sigma_8$, based on our {\it shape} method measurements), we find that $c_{2, Z_{0.47}} = 0.09 \pm 0.04$, $c_{2, Z_{0.53}} = 0.09 \pm 0.05$, and $c_{2, Z_{0.61}} = 0.09 \pm 0.03$ for $\sigma_8 = 0.789$.  The fact that there is no significant change in $c_2$ is moderately surprising and implies differences in the halo occupation distribution (HOD).  (If the HOD was not changing as a function of halo mass, $c_2$ would increase with $b_1$, see, e.g., \citealt{Nish06}).  We will discuss the HOD more in \S\ref{s:HOD}.

Of interest is the fact that the $s_3$ measurement for both $Z_{0.47}$ and $Z_{0.61}$ have a local minimum (displayed in the two left panels of Figure \ref{fig:s3}, but there is no such minimum in the $s_3$ of $Z_{0.53}$.  The minimum is at $\sim$0.6$^{\circ}$, equivalent to 11.4 $h^{-1}$Mpc for the lower redshift range, while it is at $\sim$0.3$^{\circ}$, equivalent to 8.5 $h^{-1}$Mpc, in the high redshift range. In R07, it was found that early-type galaxies also displayed a minimum in their $s_3$ measurement at $\sim$10 $h^{-1}$Mpc.  As in R07, the errors dominate the LRG measurement (though to a lesser extent), but it appears unlikely that this is a coincidence.  The feature is seen at approximately the same physical scale, but at a different angular scale due to the differences in redshift.  This rules out any possibility of observational systematics such as seeing or reddening.  It is unclear whether the feature may be due to projection effects or complicated halo dynamics that do not affect the middle redshift range.  

\section{Discussion}
We have presented a technique for measuring $\sigma_8$ using the 2 and 3-point angular area averaged correlation functions and applied it to photometrically classified LRGs from the SDSS DR5, split into three distinct redshift ranges.  Using a method that depends on the shape of $\bar{\omega}_2$ and the technique described in R07, two separate relationships were determined between $\sigma_8$ and $c_{2,LRG}$.  These relationships split the degeneracy between bias and $\sigma_8$, allowing an unambiguous determination of the 1st and 2nd order bias and $\sigma_8$.  The measured $\sigma_8$ in three redshift ranges are consistent and combine for a best-fit $\sigma_8 = 0.789 \pm 0.048$.  Our determination of $\sigma_8$ is quite precise.  It is thus important to investigate the assumptions, implicit and explicit, made when determining our measurements,  to compare our measurements to the relevant dark-matter-halo/bias theory, and to investigate how consistent our measurements are with previous results.
  
\subsection{Assumptions}
The main assumption that goes into our measurements of $c_2$ is that the bias can be expressed solely as a function of the overdensity, i.e. it is not a function of both the overdensity and the smoothing scale.  If the bias was a strong function of scale, it would invalidate any measurement made using the {\it shape} method, as a changing $b_1$ would change the shape of $\bar{\omega}_2$.  Similarly, the {\it R07} technique assumes a constant $b_1$ and $c_2$ over the range of scales that are fit.  Further, the {\it R07} measurement is fit at a different range of scales than the {\it shape} measurement.  Bias that is a strong function of scale between 2.5 $h^{-1}$Mpc and 40 $h^{-1}$Mpc would completely invalidate any comparison between measurements using the {\it shape} and the {\it R07} method, and thus invalidate our $\sigma_8$ measurements.  

Our assumption that the bias can be expressed solely as a function of the overdensity is validated by the goodness of the fit to our bias-corrected measurements.  These measurements demonstrated that, between 0.1$^\circ$ and 1.6$^\circ$, our $\bar{\omega}_2$ measurement is quite consistent with a two bias parameter model, as all of the model fits using the shape method were accepted to better than 96$\%$.  Further, in each redshift range, the probability that a single bias parameter fits the data is less than 9$\%$.  The simplest model that fits the data between 0.1$^\circ$ and 1.6$^\circ$ is thus that the bias is independent of scale and can be described by two parameters.  Therefore, we believe our measurement techniques to be valid and that our comparison of those techniques is valid as well.

The error on our $\sigma_8$ measurement is quite low in part because we held the values of $\Omega_m$, $h$, and $\Gamma$ fixed to their WMAP3 best-fit values.  Allowing these values to change does alter our best fit values of $c_2$ and $\sigma_8$.  In order to determine the degree to which the uncertainty in these parameters should affect the uncertainty of our results, we repeated our $\sigma_8$ measurements using the $Z_{0.53}$ data set ($0.5 < z < 0.57$) and produced model $\bar{\omega}_2$ and $s_3$ for each parameter at the 1$\sigma$ limits determined by the WMAP3 alone best-fits (while holding the other parameters at their best-fit values).  

As long as the geometry of the Universe is kept flat, changes in the matter density have little effect on either the amplitude or the shape of $\bar{\omega}_2$.  Thus, we expected that the value of $\Omega_m$ would have little effect on our measurement of $\sigma_8$.  This was indeed the case, as we found $\sigma_8 = 0.796 \pm 0.086$ and $0.800 \pm 0.088$ for $\Omega_m$ equaling 0.251 and 0.2134, respectively, with $\Omega_{total}$ fixed at 1 and ($\Gamma$, $h$) = (0.135, 0.73).  Conversely, we expected our measurement of $\sigma_8$ to be fairly dependent on the value of $h$, as this value significantly affects the distance to the LRGs.  We found $\sigma_8 =  0.846 \pm 0.088$ and $0.748 \pm 0.085$ for $h = 0.7$ and 0.76, respectively, with ($\Omega_{total}$, $\Gamma$, $\Omega_m$) = (1, 0.135, 0.238).  The percentage change in the measured $\sigma_8$ was approximately the same as the percentage change in $h$, suggesting a close relationship between the two (again, as expected).  

We also expected the value of $\Gamma$ to have a significant effect on the measurement, as changing $\Gamma$ alters the shape of $\bar{\omega}_2$.  For $\Gamma = 0.149$ and 0.12, we found $\sigma_8 = 0.842 \pm 0.088$ and $0.736 \pm 0.082$, with ($\Omega_{total}$, $h$, $\Omega_m$) = (1, 0.73, 0.238).  While the uncertainty in $\Gamma$ produced the largest range in $\sigma_8$ values, the percentage change in $\sigma_8$ was actually $\sim$25$\%$ smaller than the percentage change in $\Gamma$.  We thus determine that although the uncertainty in $\Gamma$ introduces the most uncertainty into our measurement of $\sigma_8$, our measurement technique is most sensitive to the value of $h$.  Despite the changes in the value of $\sigma_8$ that we measure, our measurement of $\sigma_8$ has not changed by more than 8$\%$.  This suggests that our quoted uncertainty of 0.05 would increase by less than a factor of two when uncertainties in $h$ and $\Gamma$ were taken into account.  We are thus confident our measurements are accurate to within 10$\%$.

We also assumed no error in our redshift distribution when making our measurements.  Precise knowledge of the redshift distribution is necessary for Equations 3 and 7.  In order to explicitly test our measurements' dependence on the redshift distribution, we created two new distributions for the $Z_{0.52}$ redshift range (see R07 for the details of how these distributions are constructed).  For one, we systematically increased the error of each photometric redshift by 10\% (effectively broadening the redshift distribution) and for the other we decreased the error of each photometric redshift by 10\% (effectively narrowing the distribution).  We then re-calculated the best-fit $\sigma_8$ for each distribution, finding that it increased to $0.84 \pm 0.09$ for the distribution with greater redshift errors and that it decreased to $0.76 \pm 0.09$ for the distribution with smaller redshift errors.  These differences are significant, but are smaller than our combined 1$\sigma$ error.  We thus do not believe this issue significantly adds to our quoted uncertainty.

\subsection{Testing via Halo Models}\label{s:HOD}
It is important to determine how our measured bias values compare to theoretical values.  We calculate $b_1$ and $c_2$ using halo models and the methods prescribed by \cite{Nish06}.  They show that one can combine the $N$th-order bias coefficient of halos as a function of mass and redshift, $B_N(m,z)$, the number of halos of a certain mass and redshift, $n_{halo}(m,z)$,  and the mean number of galaxies occupying a halo of a certain mass, $\left<n_g|M\right>$, to find the bias of a population of galaxies.  This can be expressed by
\begin{equation}
b_N = \frac{\int{\rm d}Mn_{halo}(M,z)B_N(M,z)\left<n_g|M\right>}{\int{\rm d}Mn_{halo}(M,z)}
\end{equation}
We determine both $n_{halo}(m,z)$ and $B_N(m,z)$, by using an ellipsoidal collapse model (e.g., \citealt{Sheth01}) and following the methods described in detail in \cite{Nish06}.  For $\left<n_g|M\right>$ we first model the number of central LRGs per halo as having a ``soft" transition between $n_g = 0$ and $n_g = 1$ such that
\begin{equation}
\left<n_{central}|M\right> = 0.5\left[1+erf\left(\frac{{\rm log}_{10}(M/M_{cut})}{\sigma_{cut}}\right)\right]
\end{equation}
as in \cite{Zheng} and \cite{Blake2}.  This helps account for the fact that we are not using a volume-limited sample, and thus at smaller redshifts the mass limit is likely to be smaller than at higher redshifts.  As in \cite{Blake2}, we model the number of satellite galaxies using a simple power law.  Thus
\begin{equation}
\left<n_{sat}|M\right> = \left(\frac{M}{M_0}\right)^{\alpha}
\end{equation}

The bias model has four free parameters, $M_{cut}$, $M_0$, $\sigma_{cut}$, and $\alpha$.  It is beyond the scope of this paper to fit for these parameters.  It is instead our intention to determine if reasonable values for these parameters can reproduce the bias values that we measure.  If this is possible, it suggests that the our bias measurements are themselves reasonable.  Using $\sigma_8 = 0.793$, $z=0.532$ and halo parameters log$_{10}(M_{cut}/M_{\odot}) = 13.3$, $\sigma_{cut} = 0.6$, log$_{10}(M_0/M_{\odot}) = 14.5$, and $\alpha = 2.0$, we find $b_1 = 1.61$ and $c_2 = 0.09$ --- results that are consistent with our best-fit values.  The values of log$_{10}(M_0/M_{\odot})$ and $\sigma_{cut}$ were chosen to be equal to the best-fit parameters found by \cite{Blake2} for LRGs with $0.5 < z < 0.55$.  The log$_{10}(M_{cut}/M_{\odot})$ and $\alpha$ parameters are slightly lower than the \cite{Blake2} parameters likely because our LRGs have a lower luminosity and thus a lower bias, minimum mass, and $\alpha$ ($\alpha$ has been seen to increase with $b_1$ in both \citealt{Blake2} and \citealt{Z05}).  This is not to suggest that we favor a steep power-law to a shallow one.  If we change $\alpha = 1.4$ (as measured by \citealt{K}), we calculate $b_1 = 1.62$ and $c_2 = 0.09$ if we also reduce log$_{10}(M_0/M_{\odot})$ to 14.41.   This implies that, from a theoretical standpoint and in the context of relevant measurements of LRG halo properties, our measurements of $b_1$ and $c_2$ are reasonable. 

The bias model can also be used to test our implicit assumption that $\langle c_2\rangle = \langle b_2\rangle/\langle b_1\rangle$.  We calculate $b_1$, $b_2$, and $c_2$ at redshifts between 0.4 and 0.7 (essentially the redshift range of our entire sample) in intervals of 0.02.  These values are used to calculate theoretical values for $\langle b_2\rangle/\langle b_1\rangle$ and $\langle b_2/b_1\rangle$.  We find that the difference between the two is less that 0.1\%, meaning that the systematic error introduced by assuming $\langle c_2\rangle = \langle b_2\rangle/\langle b_1\rangle$ is insignificant for our measurements.

\subsection{Comparison With Other Measurements}
Our measurement of $\sigma_8 = 0.789 \pm 0.050$ is consistent with most previous measurements.  There are, however, notable exceptions.  Our result is inconsistent to 1$\sigma$ with a photometric optical cluster counting technique employing SDSS data that found $\sigma_8 = 0.92 \pm 0.10$ \citep{Rozo}.  Another recent result \citep{Harker} found a similarly high value of $\sigma_8 = 0.97 \pm 0.06$ using N-body simulations in combination with semi-analytic galaxy formation models and the projected two-point correlation function of SDSS galaxies.  Other results determine $\sigma_8$ to be too small to be consistent with our measurement (to 1$\sigma$).  Notable examples are the WMAP3 constrained by inflationary models result of 0.702 $\pm 0.062$ \citep{WMAP3} and the \cite{Vod04} results that found $\sigma_8 = 0.72 \pm 0.04$ by using the cluster baryon mass function.

However, many notable results are consistent with our measurement.  The best fit WMAP3 data alone found $\sigma_8 = 0.744^{+0.05}_{-0.06}$, which is just barely consistent with our measurement to 1$\sigma$.  The WMAP3+SDSS and WMAP3+LRG best-fit values of $\sigma_8 = 0.772^{+0.036}_{-0.048}$ and $0.781^{+0.032}_{-0.045}$ are consistent with our measurement to less than 1$\sigma$.  Large disagreement with these results would be surprising, given that we used WMAP3 best-fit priors for the relevant input cosmological parameters.  The WMAP5 results are quite similar to our results, as \cite{WMAP5} find $\sigma_8 = 0.796 \pm 0.036$, for the WMAP alone five-year mean value.  More significantly, our results are also consistent to 1$\sigma$ with $\sigma_8 = 0.67^{+0.18}_{-0.13}$ as derived from optical cluster finding techniques \citep{Glad} and $\sigma_8 = 0.66 \pm 0.16$ from x-ray cluster measurements \citep{Henry}.  Due to the fact that the inconsistent measurements appear to be as likely to be lower than our measurement as they are higher, we feel that these results are hinting at a potential convergence to a $\sigma_8$ value that is close to 0.8.

\section{Conclusions}
We present and test a new method for determining the value of $\sigma_8$.  The method and the results of our testing can be summarized as follows:

$\bullet$ The technique for measuring $\sigma_8$ utilizes two measures of the second-order bias of a density field.  The two methods have different dependencies on the value of $\sigma_8$, and can thus be combined to determine a best-fit $\sigma_8$ and second-order bias.  One measure of the second-order bias (the $R07$ method) has been used many times before (e.g., R07) and depends on the amplitude of $s_3$.  The other method (the $shape$ method) has (to our knowledge) never been used before.  It relies on correcting the overdensities for given first- and second-order bias parameters and determining the bias parameters that allow the shapes of the $\bar{\omega}_2$ measurement and model to become consistent.

$\bullet$ The method was tested using a mock catalog of LRGs drawn from the Blaizot all-sky catalog \citep{blaiz} that was constructed using Millennium Simulation data \citep{mill} and its input cosmological parameters for $\Omega_m$, $h$, and $\Gamma$.  The Millennium Simulation assumed $\sigma_8 = 0.9$ and we measured $\sigma_8 = 0.898 \pm 0.062$.   This measurement proved that our method is both accurate and precise. 

$\bullet$ We photometrically selected LRGs from SDSS DR5, determined photometric redshifts for each LRG, and removed stars following the prescriptions of \cite{2slaq}.  We divided this LRG catalog into three samples by redshift, with the separate ranges being $0.4 < z < 0.5$ ($Z_{0.47}$), $0.5 < z < 0.57$ ($Z_{0.53}$), and $0.57 < z < 0.7$ ($Z_{0.61}$). 

$\bullet$ We measured $\sigma_8$ in each sample and found $\sigma_8 = 0.776 \pm 0.080$, $\sigma_8 = 0.796 \pm 0.086$, and $\sigma_8 = 0.798 \pm 0.094$, respectively.  Combining these consistent results we determined $\sigma_8 = 0.789 \pm 0.050$.  For each measurement, we assumed the relevant cosmological parameters were equal to their WMAP3 best-fit values.  

$\bullet$ Allowing the relevant cosmological parameters to vary within their WMAP3 1$\sigma$ error, we found our measurement of $\sigma_8$ changed by less than 8$\%$.  Thus, even allowing for these uncertainties, our method produces a precise measurement that we are confident is accurate to within 10$\%$.

$\bullet$ We measured $b_{1,Z_{0.47}} = 1.47 \pm 0.09$, $b_{1,Z_{0.53}} = 1.65 \pm 0.09$, and $b_{1,Z_{0.61}} = 1.80 \pm 0.10$ and $c_{2, Z_{0.47}} = 0.09 \pm 0.04$, $c_{2, Z_{0.53}} = 0.09 \pm 0.05$, and $c_{2, Z_{0.61}} = 0.09 \pm 0.03$.  Using a halo model, we determined that the bias values for $Z_{0.53}$ were consistent with reasonable, and previously measured, HOD parameters. 

The techniques described herein can easily be repeated and tested using other cosmic samples.  Future surveys, such as the Panoramic Survey Telescope $\&$ Rapid Response System (Pan-STARRS), the Dark Energy Survey (DES), and the Large Synoptic Survey Telescope (LSST), will provide opportunities to measure both the 2- and 3-point area averaged correlation functions to extreme precision.  Repeating the techniques we have presented here will allow researchers to determine $\sigma_8$ accurately, precisely, and independently of future cluster counting and CMB techniques, thus providing an important check on those results.  

In addition to our determinations of the bias and of $\sigma_8$, we have found that a feature in the hierarchical amplitude of $s_3$ at $\sim$10 $h^{-1}$Mpc exists in two of the three data sets.  Given that a feature existed at approximately the same physical location for early-type galaxies at significantly smaller redshifts (R07), the feature appears physical in nature.  The fact that it is absent in one of our data sets hints that the possible cause may be due to projection effects, or that perhaps the feature is indicative of complicated halo occupation statistics.  The feature demands further study, both observationally and theoretically, and we are currently focusing our efforts to explain this phenomon.

\acknowledgements

AJR, RJB and ADM acknowledge support from Microsoft Research, the University of Illinois, and NASA through grant NNG06GH156.  The authors made extensive use of the storage and computing facilities at the National Center for Supercomputing Applications and thank the technical staff for their assistance in enabling this work.

We thank Ani Thakar and Jan Van den Berg for help with obtaining a copy of the SDSS DR5 databases. 
We thank Ravi Sheth for comments that helped improve the paper.  We thank an anonymous referee whose comments helped us to significantly improve our results.

Funding for the creation and distribution of the SDSS Archive has been provided by the Alfred P. Sloan Foundation, the Participating Institutions, the National Aeronautics and Space Administration, the National Science Foundation, the U.S. Department of Energy, the Japanese Monbukagakusho, and the Max Planck Society. The SDSS Web site is http://www.sdss.org/.

The SDSS is managed by the Astrophysical Research Consortium (ARC) for the Participating Institutions. The Participating Institutions are The University of Chicago, Fermilab, the Institute for Advanced Study, the Japan Participation Group, The Johns Hopkins University, the Korean Scientist Group, Los Alamos National Laboratory, the Max-Planck-Institute for Astronomy (MPIA), the Max-Planck-Institute for Astrophysics (MPA), New Mexico State University, University of Pittsburgh, University of Portsmouth, Princeton University, the United States Naval Observatory, and the University of Washington.

\clearpage

\clearpage

\clearpage

 \end{document}